\RequirePackage{fix-cm}
\documentclass[smallextended]{svjour3}       
\smartqed  
\usepackage{graphicx}
\usepackage{subfigure}
\usepackage{amssymb}
\usepackage{color}
\usepackage{amsmath}
\usepackage{cite}
\usepackage{graphicx}
\usepackage{epstopdf}
\usepackage{cases}
\begin{document}

\title{Generating mechanism and dynamic of the smooth positons for the derivative nonlinear Schr\"odinger equation
 \thanks{*Corresponding author, Maohua Li: limaohua@nbu.edu.cn}
}
\subtitle{}


\author{Wenjuan Song      \and  Shuwei Xu   \\   \and
        Maohua Li*   \and  Jingsong He
}


\institute{Wenjuan Song  \at
              School of Mathematics and Statistics, Ningbo University, Zhejiang 315211, P.\ R.\ China
              \and
              Shuwei Xu \at
               College of Mathematics Physics and Information Engineering, Jiaxing University,
                  Jiaxing, Zhejiang 314001, P.\ R.\ China
               \and
              Maohua Li \at
              School of Mathematics and Statistics, Ningbo University, Zhejiang 315211, P.\ R.\ China\\
              \email{limaohua@nbu.edu.cn}
               \and
           Jingsong He \at
              Institute for Advanced Study, Shenzhen University, Shenzhen, Guangdong 518060, P.\ R.\ China
      \\
}

\date{Received: date / Accepted: date}

\maketitle

\begin{abstract}
Based on the degenerate Darboux transformation, the $n$-order smooth positon solutions for the derivative nonlinear Schr\"{o}dinger equation are generated by means of the general determinant expression of the $N$-soliton solution, and interesting dynamic behaviors of the smooth positons are shown by the corresponding three dimensional plots in this paper. Furthermore, the decomposition process, bent trajectory and the change of the phase shift for the positon solutions are discussed in detail. Additional, three kinds of mixed solutions, namely (1) the hybrid of one-positon and two-positon solutions, (2) the hybrid of two-positon and two-positon solutions, and (3) the hybrid of one-soliton and three-positon solutions are presented and their rather complicated dynamics are revealed.
\end{abstract}

\section{Introduction}

As a degenerate cases of Wronskian formula of solutions of the Korteweg-de Vries (KdV) equation,  Matveev \cite{Matveev19992(1)} firstly introduced the positons of the KdV equation which is a singular new solution of nonlinear evolution equations. Matveev also obtained positon and soliton-positon solutions of the the KdV equation and summarized many significant properties of the positon solutions which is different from the soliton
solutions. There exists a class of slowly decreasing reflectionless potential which is called  potentials superreflectionless \cite{Matveev19992}. Compared with the soliton solutions with exponential decay, the positons were weakly localized. There are two reason for the name called positon solutions that is generated by a positive spectral singularity embedded in the spectrum and always positive in a small enough neighborhood of the pole. About the other interacting objects, the positons are completely transparent. In particular, two positons remain unchanged after mutual collision while during the positon and soliton collision, the soliton solution remain unchanged, however, the positon is affected of which the carrier wave and envelope produce the finite phase shift \cite{Chow1998,Dubard2010}. In Ref.\cite{Matveev1994}, the authors established the connection of the solutions of the positons, soliton and breather. Inspiringly by the pioneering work of Matveev, such solutions have been extended to other well-known equations, such as the defocusing modified KdV (mKdV) equation \cite{positon1992,positon1995}, the sine-Gordon (sG) equation \cite{Beutler1993}, the Toda lattice \cite{Stahlofen1995}, the Hirota-Satsuma coupled KdV system \cite{hu2008}. It is crucial to get the smooth positons because the positon solutions above-mentioned are singular. Recently, the smooth positons of focusing mKdV equation \cite{xingqiuxia2017}, complex mKdV equation \cite{liuwei2017} and the second-type derivative nonlinear Schr\"odinger (DNLSII) equation \cite{liushuzhi2019} also have been constructed. Postion is a slowly decreasing analogue of soliton, which is closed related to Wigner-von Neumann phenomenon \cite{tmp2002}. It is known that the positon of the KdV, as one kind of potential in quantum mechanics, is expected to be realized \cite{tmp2002} in practice using band engineering technique \cite{Nature1992}. This interesting application of the positon inspires us to study the positon in other soliton equation with strong physical background.

 The nonlinear Schr\"odinger (NLS) equation is one of the most significant equations in physics and mathematics and it can be derived from the Ablowitz-Kaup-Newell-Segular \cite{AMJ1973,ZS1972}. Various modifications of the equation have been investigated extensively and discussed intensively. One of these attempts is to study the effects of higher order perturbations which has been proposed by various authors. Considering higher order nonlinear effect, the derivative nonlinear Schr\"odinger (DNLS) equations with a polynomial spectral problem of arbitrary order \cite{KBG1981} are regarded as the models in a wide variety of fields such as weekly nonlinear dispersive water waves \cite{RSJ1977}. The DNLS equations have three generic deformations, i.e. the DNLSI equation \cite{ME1976}
 \begin{equation}\label{1}
iq_t-q_{xx}+i(q^2q^*)_x=0,
\end{equation}
\noindent the DNLSII equation \cite{CLL1979}, which is also called the Chen-Lee-Liu (CLL) equation, of the form
 \begin{equation}\label{2}
iq_t+q_{xx}+iqq^*q_x=0,
\end{equation}
\noindent and the DNLSIII equation (or the GI equation) \cite{GI1983}
 \begin{equation}\label{3}
iq_t+q_{xx}-iq^2q^{*}_x+\frac{1}{2}q^3q^{*2}=0.
\end{equation}

The equation (\ref{1}) is also briefly called the DNLS equation, which is one of the most important integrable systems in the mathematics and physics, where $*$ is the complex conjugation, and the subscript $x$ (or $t$) denotes the partial derivative with respect to $x$ (or $t$). The DNLS equation is connected with the DNLSII equation by a simple gauge transformation \cite{Wadati1983,Kundu1984} and the relationship of DNLS equation and DNLSIII equation is also discussed in Ref.\cite{Kakei1995}.

The DNLS equation was proposed to describe Alfv\'{e}n waves in plasma that is the wave of finite amplitude which propagate parallel to the magnetic field \cite{ME1976,Spangler1997}. On the one hand, the equation is used to describe large-amplitude magnetohydrodynamic (MHD) wave propagating in plasmas at moderate angles with respect to the equilibrium magnetic field \cite{Fedum2008}. Truncated DNLS equation \cite{SAG2007}and DNLS with nonlinear Landau dumpling \cite{SAG2010} are put into use in practical plasmas as well. More importantly, the equation is applied to nonlinear optics, such as, the sub-picosecond or femtosecond pluses in single-mode optical fibers \cite{Tzoar1981,Anderson1983,Govind2001}.

The DNLS equation is one of the rare several integrable nonlinear models that permit soliton solutions. Under vanishing boundary condition (VBC), Kaup and Newell \cite{KN1978} solved the appropriate inverse scattering problem and obtained the one soliton solution. By using inverse scattering method, the soliton are examined analytically and numerically under vanishing boundary condition and nonvanishing boundary condition (NVBC), introduce the ``paired soliton" which generally pulsates with a period but degenerate to a stable ``pure soliton" (bright and dark solitons) in a limited case \cite{Kawata1979}. An explicit expression for the $N$-soliton solution is expressed in terms of determinants by means of algebraic techniques that solving the reduced Zakharov-Shabat equations \cite{HNN1990}.

Furthermore, in order to avoid constructing Riemann sheets, an inverse scattering transform (IST) for the DNLS equation with NVBC is derived by introducing an affine parameter. A one-soliton solution which is a breather and degenerates to a bright or dark soliton as the discrete eigenvalue becomes purely imaginary \cite{Chen2004}, simpler than that in the literature \cite{ME1989}, is obtained. And the bilinearization of a generalized derivative nonlinear schr\"{o}dinger equation is also discussed  and the solitons solution are constructed as the quotients of the Wronski-type determinants \cite{Kakei1995}.

Recently, $N$-soliton solution of two component DNLS equation is investigated by the two-fold Darboux transformation (DT) \cite{Ling2010}. And Chan \cite{CHN2014} present the rogue waves of DNLS equation using a long-wave limit of breather. Later, the rogue wave of coupled DNLS equation by means of low-frequency limit of breather is derived \cite{CHN2016}.
In Ref.\cite{xushuwei2011}, the determinant representation of the $n$-fold DT and formulae of $q^{[n]}$ and $r^{[n]}$ is expressed. The complete classification of the dark soliton, bright soliton as well as periodic solution are given and obtained rational traveling solution and rogue wave. As for mixed nonlinear Schr\"{o}dinger (MNLS) equation, which an integrable equation with the nonlinear term in NLS equation and DNLS equation denote the effects of phasemodulation (SPM) and self-steepening from the points of view mathematics and physics, the rational solutions is also investigated in Ref.\cite{He2015}. Those results have also been extended to other NLS-type equation with additional derivative terms \cite{Qiu2015,Yuan2017}. It is worthwhile and natural to know whether or not it has other solutions or interesting dynamics. Inspired by the result above, the smooth positon solutions of DNLS equation and it's dynamic behaviors are also worthwhile studied. To the best of our knowledge, the smooth positon solutions of DNLS equation have never been reported. The main aim of this paper is to obtain the multi-positon solution of the DNLS equation and studied its property of dynamics.

This paper is organized as follows. In section \ref{section2}, the explicit formula of smooth positons is obtained by means of DT and degenerate DT of the DNLS equation. In  section \ref{section3}, with the special higher Taylor expansion, the positons are decomposed by modulus square, and then also discussed the trajectory and ``phase shift" in detail. In section \ref{section4}, the combinations of solitons and positons are investigated. The conclusion is provided in  section \ref{section5}.

\section{Positons of DNLS equation}\label{section2}
The couple of the derivative nonlinear Schr\"{o}dinger equations \cite{KN1978},
\begin{equation}\label{4}
q_t+iq_{xx}-(rq^2)_x=0,
\end{equation}
\begin{equation}\label{5}
r_t-ir_{xx}-(r^2q)_x=0,
\end{equation}
are exactly lead to the DNLS equation for $r=-q^*$  but the choice $r=q^*$ would result in equation (\ref{1}) with the sign of the nonlinear term changed. The $*$ denotes the complex conjugation.

The Lax pairs of the coupled DNLS equations (\ref{4}) and (\ref{5}) can be derived by the KN spectral problem \cite{KN1978}
\begin{equation}\label{6}
\partial_x\psi=(J\lambda^2+Q\lambda)\psi=U\psi,
\end{equation}
\begin{equation}\label{7}
\partial_t\psi=(2J\lambda^4+V_3\lambda^3+V_2\lambda^2+V_1\lambda)\psi=V\psi,
\end{equation}
with
$$
\psi=\left(\begin{array}{ccc}
\phi  \\
\varphi \\
\end{array}\right),
J=\left(\begin{array}{ccc}
i & 0 \\
0 & -i \\
\end{array}\right),
Q=\left(\begin{array}{ccc}
0 & q \\
r & 0 \\
\end{array}\right),
$$

$$
V_3=2Q,
V_2=Jqr,
V_1=\left(\begin{array}{ccc}
0 & -iq_x+q^2r \\
ir_x+r^2q & 0 \\
\end{array}\right).
$$
Here $\lambda$ (an arbitrary complex number) is called the eigenvalue or spectral parameter and $\psi$ is the eigenfunction associated with $\lambda$ of the KN system. The zero curvature equation $U_t-V_x+[U,V]=0$ infers the couple of the DNLS equations (\ref{4}) and (\ref{5}).

The determinant representation of the $N$-fold DT for the DNLS equation is given in Ref.\cite{xushuwei2011} which is similar to the determinant representation of the $N$-fold DT for the NLS equation \cite{He2006}. The  formulas of the soliton and positon of DNLS equation is obtained explicitly by setting seed solution $q=0$,  and the eigenfunction
\begin{equation}\label{8}
\psi_j=\psi(\lambda_j)=\left(\begin{array}{ccc}
\phi_j(\lambda_j) \\
\varphi_j(\lambda_j) \\
\end{array}\right)=\left(\begin{array}{ccc}
\exp{(i(\lambda_j^2x+2\lambda_j^4t))}\\
\exp{(-i(\lambda_j^2x+2\lambda_j^4t)} \\
\end{array}\right)
\end{equation}
is associated with eigenvalue $\lambda_j=\alpha_j+i\beta_j$ in Theorem 2 of Ref.\cite{xushuwei2011}, then an explicit form of the $N$-soliton of the DNLS equation is:
\begin{equation}\label{9}
q^{[n]}=\frac{\Omega_{11}^2}{\Omega_{21}^2}q+2i\frac{\Omega_{11}\Omega_{12}}{\Omega_{21}^2},
r^{[n]}=\frac{\Omega_{21}^2}{\Omega_{11}^2}r-2i\frac{\Omega_{21}\Omega_{22}}{\Omega_{11}^2}.
\end{equation}
where
$$
{\Omega_{11}}=\left|\begin{array}{ccccccc}
\lambda_{1}^{2n-1}\varphi_{1} &  \lambda_1^{2n-2}\phi_{1} &\lambda_{1}^{2n-3}\varphi_{1}&   \cdots\lambda_{1}\varphi_{1}&  \phi_{1}\\
\lambda_{2}^{2n-1}\varphi_{2} &  \lambda_2^{2n-2}\phi_{2} &\lambda_{2}^{2n-3}\varphi_{2}&   \cdots\lambda_{2}\varphi_{2}&  \phi_{2}\\
\lambda_{3}^{2n-1}\varphi_{3} &  \lambda_3^{2n-2}\phi_{3} &\lambda_{3}^{2n-3}\varphi_{3}&   \cdots\lambda_{3}\varphi_{3}&  \phi_{3}\\
\vdots                       &            \vdots        &\vdots                      &    \vdots                     &  \vdots\\
\lambda_{2n}^{2n-1}\varphi_{2n} &  \lambda_{2n}^{2n-2}\phi_{2n} &\lambda_{2n}^{2n-3}\varphi_{2n}&   \cdots\lambda_{2n}\varphi_{2n}&  \phi_{2n}\\
\end{array}\right|,
$$

$$
{\Omega_{12}}=\left|\begin{array}{ccccccc}
\lambda_{1}^{2n}\phi_{1}  &  \lambda_1^{2n-2}\phi_{1} &\lambda_{1}^{2n-3}\varphi_{1}&   \cdots\lambda_{1}\varphi_{1}&  \phi_{1}\\
\lambda_{2}^{2n}\phi_{2}  &  \lambda_2^{2n-2}\phi_{2} &\lambda_{2}^{2n-3}\varphi_{2}&   \cdots\lambda_{2}\varphi_{2}&  \phi_{2}\\
\lambda_{3}^{2n}\phi_{2}  &  \lambda_3^{2n-2}\phi_{2} &\lambda_{3}^{2n-3}\varphi_{2}&   \cdots\lambda_{3}\varphi_{3}&  \phi_{3}\\
\vdots                   &            \vdots        &\vdots                      &    \vdots                     &  \vdots\\
\lambda_{2n}^{2n}\phi_{2n}  &  \lambda_{2n}^{2n-2}\phi_{2n} &\lambda_{2n}^{2n-3}\varphi_{2n}&   \cdots\lambda_{2n}\varphi_{2n}&  \phi_{2n}\\
\end{array}\right|,
$$

$$
{\Omega_{21}}=\left|\begin{array}{ccccccc}
\lambda_{1}^{2n-1}\phi_{1} &  \lambda_1^{2n-2}\varphi_{1} &\lambda_{1}^{2n-3}\phi_{1}&   \cdots\lambda_{1}\phi_{1}&  \varphi_{1}\\
\lambda_{2}^{2n-1}\phi_{2} &  \lambda_2^{2n-2}\varphi_{2} &\lambda_{2}^{2n-3}\phi_{2}&   \cdots\lambda_{2}\phi_{2}&  \varphi_{2}\\
\lambda_{3}^{2n-1}\phi_{3} &  \lambda_2^{2n-2}\varphi_{3} &\lambda_{2}^{2n-3}\phi_{3}&   \cdots\lambda_{2}\phi_{3}&  \varphi_{3}\\
\vdots                    &            \vdots           &\vdots                   &    \vdots                  &  \vdots\\
\lambda_{2n}^{2n-1}\phi_{2n} &  \lambda_{2n}^{2n-2}\varphi_{2n} &\lambda_{2n}^{2n-3}\phi_{2n}&   \cdots\lambda_{2n}\phi_{2n}&  \varphi_{2n}\\
\end{array}\right|,
$$

$$
{\Omega_{22}}=\left|\begin{array}{ccccccc}
\lambda_{1}^{2n}\varphi_{1} &  \lambda_1^{2n-2}\varphi_{1} &\lambda_{1}^{2n-3}\phi_{1}&   \cdots\lambda_{1}\phi_{1}&  \varphi_{1}\\
\lambda_{2}^{2n}\varphi_{2} &  \lambda_2^{2n-2}\varphi_{2} &\lambda_{2}^{2n-3}\phi_{2}&   \cdots\lambda_{2}\phi_{2}&  \varphi_{2}\\
\lambda_{3}^{2n}\varphi_{3} &  \lambda_3^{2n-2}\varphi_{3} &\lambda_{3}^{2n-3}\phi_{3}&   \cdots\lambda_{3}\phi_{3}&  \varphi_{3}\\
\vdots                     &            \vdots           &\vdots                   &    \vdots                  &  \vdots\\
\lambda_{2n}^{2n}\varphi_{2n} &  \lambda_{2n}^{2n-2}\varphi_{2n} &\lambda_{2n}^{2n-3}\phi_{2n}&   \cdots\lambda_{2n}\phi_{2n}&  \varphi_{2n}\\
\end{array}\right|.
$$

\noindent When $n=1$, the explicit formula of one-soliton solution is following:
\begin{equation}\label{q1-s}
q_{1-s}=4i\alpha_1\beta_1\frac{(-i\alpha_1\cosh(4\alpha_1\beta_1H)+\beta_1\sinh(4\alpha_1\beta_1H))^3}{((-\alpha_1^2-\beta_1^2)\cosh(4\alpha_1\beta_1H)^2+\beta_1^2)^2}\exp(2ih),
\end{equation}
where $H=4\alpha_1^2t-4\beta_1^2t+x$, $h=-\beta_1^2x+2\beta_1^4t-12\alpha_1^2\beta_1^2t+\alpha_1^2x+2\alpha_1^4t$.

When $\lambda_1=\lambda_3$, the denominator of two-soliton is zero by choosing $n=2$ in eq.(\ref{9}). In general, we get degenerate $N$-fold DT for DNLS equation  by setting $\lambda_{2j-1}=\lambda_1+\epsilon$ ($j = 2,3\ldots,n$ ) , and then the solution $q^{[n]}$ also becomes an indeterminate form $\frac{0}{0}$. In the following, we can get the smooth positon solutions of the DNLS equation using the degenerate DT and the higher-order Taylor expansion with $\lambda_{2j-1}=\lambda_1+\epsilon$ ($j = 2,3\ldots,n $) which is similar to construct the degenerate solitons from a zero seed of mKdV equation. Substituting $\psi_j$ into this degenerate $N$-fold DT, and setting $N_{2n}$ and $W_{2n}$ as the following forms, the $n$-positon of the DNLS equation is constructed.

\noindent \textbf{Proposition 1}: The $N$-soliton solution with a ``seed" solution $q=0$ from $N$-fold DT in the degenerate limit $\lambda_{2j-1}\rightarrow\lambda_1$ generates a $n$-positon solution of DNLS equation, which is given by
\begin{equation}\label{npositon}
q_{n-p}=2i\frac{N_{2n}'}{W_{2n}'},
\end{equation}
where$${N_{2n}}={\Omega_{11}}{\Omega_{12}}$$
$${W_{2n}}={\Omega_{21}}{\Omega_{21}}$$

$$
N_{2n}'=(\frac{\partial^{n_{i}-1}}{\partial\epsilon^{n_{i}-1}}|_{\epsilon=0}(N_{2n})_{ij}(\lambda_{1}+\epsilon))_{2n\times2n},
$$
$$
W_{2n}'=(\frac{\partial^{n_{i}-1}}{\partial\epsilon^{n_{i}-1}}|_{\epsilon=0}(W_{2n})_{ij}(\lambda_{1}+\epsilon))_{2n\times2n},
$$
and $n_{i}=[\frac{i+1}{2}]$, ${[i]}$ define the floor function of $i$.

In above proposition, the reduction conditions are $\lambda_{2j}=-\lambda_{2j-1}^*$, $\phi_{2j}=\varphi_{2j-1}^*$, $\varphi_{2j}=\phi_{2j-1}^*$, $(j=1,2,3,...,n)$ .

we only present the explicit form of the two-positon by letting $n=2$ in Proposition 1 because of the tedious mathematical formulas. The explicit expression of two-positon solution is following:
\begin{equation}\label{q2-p}
q_{2-p}=\frac{A_1A_2}{(512\omega_3-512\omega_4-(\alpha_1^2+\beta_1^2)(\omega_5\cosh(8\alpha_1\beta_1H)+\omega_6\sinh(8\alpha_1\beta_1H)))^2},
\end{equation}\\
where
$H=4\alpha_1^2t-4\beta_1^2t+x$,$h=-\beta_1^2x+2\beta_1^4t-12\alpha_1^2\beta_1^2t+\alpha_1^2x+2\alpha_1^4t$,\\

$A_1=256I\alpha_1\beta_1\exp(2ih)(\omega_1\cosh(4\alpha_1\beta_1H)+\omega_2\sinh(4\alpha_1\beta_1H))$,\\

$A_2=((\alpha_1^2+\beta_1^2)(\omega_5\cosh(8\alpha_1\beta_1H)-\omega_6\sinh(8\alpha_1\beta_1H))-512\omega_3-512\omega_4)$,\\

$\omega_1=(3-3i)\beta_1^2\alpha_1^5t+\frac{1}{16}(1+i)+(2-2i)\beta_1^4t+(1/4)(1-i)\beta_1^2x)\alpha_1^3+(-1+i)(\beta_1^2t-\frac{x}{4})\beta_1^4\alpha_1$,\\

$\omega_2=(-1-i)\beta_1\alpha_1^6t+(2+2i)(\beta_1^2t-\frac{x}{8})\beta_1\alpha_1^4+(3+3i)(\beta_1^2t-\frac{x}{12}))\beta_1^3\alpha_1^2+\frac{1}{16}(-1+i)\beta_1^3$,\\

$\omega_3=\omega_{31}+\omega_{32}+\omega_{33}+\omega_{34}$,\\

$\omega_{31}=(-1+i)\beta_1^2\alpha_1^{10}t^2+(-4+4i)(\beta_1^2t+\frac{x}{8})\beta_1^2\alpha_1^8t$,\\

$\omega_{32}=6((-1+i)\beta_1^4t^2+\frac{1}{12}(-1+i)\beta_1^2xt+\frac{1}{96}((-1+i)x^2))\beta_1^2\alpha_1^6$,\\

$\omega_{33}=(\frac{1}{512}(-1+i)+(-4+4i)\beta_1^8t^2+(1/2)(1-i)\beta_1^6xt+\frac{1}{8}(-1+i)\beta_1^4x^2)\alpha_1^4$,\\

$\omega_{34}=(\beta_1^2t-\frac{x}{4})\beta_1^4((-1+i)\beta_1^4t+\frac{1}{4}(1-i)\beta_1^2x)\alpha_1^2+\frac{1}{512}(-1+i)\beta_1^4$,\\

$\omega_4=-\frac{1}{8}(1+i)\alpha_1^2(\alpha_1^4t+(-6\beta_1^2t+(1/4)x)\alpha_1^2+\beta_1^2(\beta_1^2t-\frac{1}{4}x))\beta_1^2$,\\

$\omega_5=(1-i)(\alpha_1^2-\beta_1^2)$,\\

$\omega_6=-(2+2i)\alpha_1\beta_1$.\\

In Proposition 1, $n$-positon generated by the degenerated DT and the higher-order Taylor expansion is smooth and expressed by mixture of exponential functions and polynomials of $x$ and $t$, which is different substantially from the soliton expressed by exponential functions and the rogue wave represented by the polynomials.

We provide the three-dimensional evolutions of two-positon (Fig.\ref{1}(a)), three-positon (Fig.\ref{2}(a)), four-positon (Fig.\ref{3}(a)). In order to see clearly their trajectories, the density plots of positons are given (Fig.\ref{1}(b)), (Fig.\ref{2}(b)), (Fig.\ref{3}(b)), respectively. The trajectory  will be discussed in the following section in terms of exact and approximate ways.

\section{Dynamics of the positons of DNLS}\label{section3}

The dynamical properties of the positon solution of DNLS equation will be discussed in this section. It is clearly that the $q_{2-p}$ is not a traveling wave with a constant profile from the formula of the solution and the trajectory of the two-positon is a slowly changing curve instead of a straight line. By means of the higher Taylor expansion in Proposition 1, the trajectory, decomposition, and the ``phase shift" of the positons are introduced in the following Proposition and we study the evolution of the positon solution of the DNLS equation with the decomposition of the modulus square.

\noindent \textbf{Proposition 2}: As $|t|\rightarrow\infty$, the modulus square of a two-positon solution of the DNLS equation is decomposed as following form:

 \begin{equation}\label{13}
 |q_{2-p}|^2\approx |q_{1-s}(H+\frac{\ln(t^4)}{16\alpha_1\beta_1})|^2+|q_{1-s}(H-\frac{\ln(t^4)}{16\alpha_1\beta_1})|^2
 \end{equation}

\noindent a more precise approximate trajectory are two curves defined by

\noindent $H\pm\frac{\ln(8388608\alpha_1^8\beta_1^8(\alpha_1^2+\beta_1^2)^2t^4)}{16\alpha_1\beta_1}=0$, where $H=4\alpha_1^2t-4\beta_1^2t+x$.\\

\noindent \textbf{Proof}: It is well-known that the $N$-soliton solution can be decomposed into $N$ single soliton with the ``phase shift" when $|t|>>0$.
 This fact stimulates strongly us to consider a similar decomposition about the multi-positon because the multi-positon is the degenerate limit of a multi-soliton. In this proposition, we start from two-positon solution in terms of the decomposition of the modulus square, i.e.
\begin{equation}\label{14}
{|q_{2-p}|^2}\approx |q_{1-s}(H+c_{11})|^2+|q_{1-s}(H-c_{11})|^2
\end{equation}\\
when $|t|\rightarrow\infty$, and in which

\begin{equation}\label{15}
q_{1-s}(\theta)=4i\alpha_1\beta_1\frac{(-i\alpha_1\cosh(4\alpha_1\beta_1\theta)+\beta_1\sinh(4\alpha_1\beta_1\theta))^3}{((-\alpha_1^2-\beta_1^2)(\cosh(4\alpha_1\beta_1\theta)^2+\beta_1^2)^2}\exp(2ih),
\end{equation}\\
where $\theta=H\pm{c_{11}}$.

It is easy to see that Eq (\ref{15}) is a one-soliton in equation (\ref{q1-s}) with a ``phase shift" $c_{11}$  and worth mentioning that the ``phase shift" for a usual two-soliton is a constant, but ``phase shift" $c_{11}$ is the undetermined function of $x$ and $t$ which will be given later. In order to get the ``phase shift" $c_{11}$,  substitute (\ref{15}) into (\ref{14}) by a simple calculation, and just consider the corresponding approximation of this equation in the neighborhood of $H=0$ when $|t|\rightarrow\infty$, then it yields
\begin{equation}\label{16}
-16777216\alpha_1^8\beta_1^8(\alpha_1^2+\beta_1^2)^4t^4+e^{16\alpha_1\beta_1c_1}+e^{-16\alpha_1\beta_1c_1}-16\alpha_1^8+6\alpha_1^4-4\alpha_1^2\beta_1^2+6\beta_1^4\approx0.
\end{equation}\\

Solving above equation, then $c_{11}\approx\frac{ln(8388608\alpha_1^8\beta_1^8(\alpha_1^2+\beta_1^2)^2t^4)}{16\alpha_1\beta_1}$. So equation (\ref{14}) also holds when $c_{11}$ is replaced by a simplified form of approximation $c_1=\frac{\ln(t^4)}{16\alpha_1\beta_1}$ as $|t|\rightarrow\infty$, which implies equation (\ref{13}).\\

\noindent \textbf{Remark 3.1} : In proposition 2, the ``phase shift" of positon $c_{11}$ is a function of $t$ which is different from the ``phase shift" of soliton solution that is usually a constant. It's connected with $\ln{t^4}$, differing from the focusing mKdV equation (see Ref.\cite{xingqiuxia2017}) and the complex mKdV equation (see Ref.\cite{liuwei2017}). The ``phase shift" of the focusing mKdV equation  and the complex mKdV equation are related with the $\ln{t^2}$.\\
\begin{figure}[!htbp]
\centering
\subfigure[]{\includegraphics[height=4.2cm,width=4.2cm]{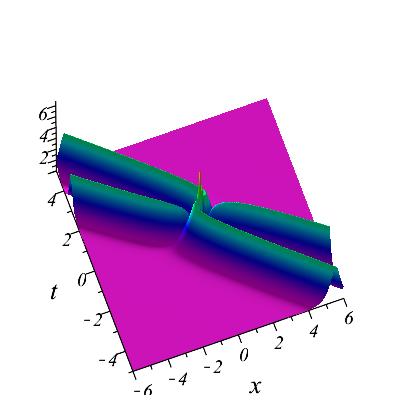}}
\subfigure[]{\includegraphics[height=4.2cm,width=4.2cm]{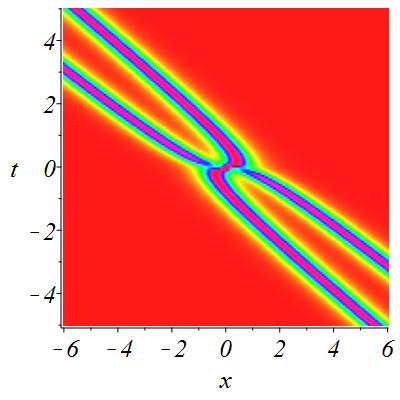}}
\subfigure[]{\includegraphics[height=4.2cm,width=4.2cm]{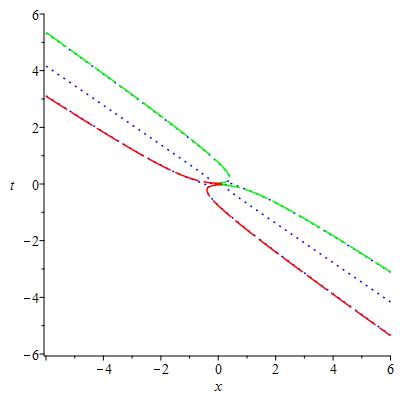}}
\caption{The evolution of two-positon $|q_{2-p}|^2$ with the parameter $\alpha_1=1$ and $\beta_1=0.8$ of the DNLS equation on $(x,t)$-plane. (a) is 3D plot, (b) is the density plot, (c) is the trajectory and the decomposition of two-positon solution with $c_{11}$. Three blue dots line in panel (c) for extreme values, two maxima denote the trajectories, but one minima (middle) is not. Two dashed line (green, upper; red, lower) denote approximate trajectories.}\label{fig1}
\end{figure}

\noindent \textbf{Proposition 3}: As $|t|\rightarrow\infty$, the modulus square of a three-positon solution of the DNLS equation is decomposed as following form:

 \begin{equation}\label{17}
 |q_{3-p}|^2\approx |q_{1-s}(H+\frac{\ln(t^{16})}{16\alpha_1\beta_1})|^2+|q_{1-s}(H)|^2+|q_{1-s}(H-\frac{\ln(t^{16})}{16\alpha_1\beta_1})|^2
 \end{equation}\\

\noindent a more precise approximate trajectory are three curves defined by

\noindent $H\pm\frac{ln(\frac{\beta_1^{32}\alpha_1^{36}(\alpha_1^2+\beta_1^2)^{14}t^{16})}{2}}{16\alpha_1\beta_1}=0$ and $H=0$ ($H=4\alpha_1^2t-4\beta_1^2t+x$).\\

\noindent \textbf{Proof}: Similar to proposition 2, we take a similar decomposition of three-positon solution into account, i.e. suppose\\
\begin{equation}\label{18}
 |q_{3-p}|^2\approx |q_{1-s}(H+c_{22})|^2+|q_{1-s}(H)|^2+|q_{1-s}(H-c_{22})|^2
 \end{equation}

\noindent is correct when $|t|\rightarrow\infty$. Certainly, ``phase shift" $c_{22}$ is an undetermined function of function $x$ and $t$ as before, and $q_{1-s}(H\pm{c_{22}})$ is obtained by a simple replacement of $c_{11}$ in equation (\ref{15}). Substitute $q_{1-s}(H+c_{22})$, $q_{1-s}(H)$ and $q_{1-s}(H+c_{22})$ in equation(\ref{18}), consider the corresponding approximation in the neighborhood of $H=0$, then it yields
\begin{equation}\label{19}
e^{(16\alpha_1\beta_1c_2)}+e^{(-16\alpha_1\beta_1c_2)}-\beta_1^{32}\alpha_1^{36}(\alpha_1^2+\beta_1^2)^{14}t^{16}\approx0
\end{equation}

\noindent After that, $c_{22}\approx\frac{ln(\frac{\beta_1^{32}\alpha_1^{36}(\alpha_1^2+\beta_1^2)^{14}t^{16})}{2}}{16\alpha_1\beta_1}$ is an approximate solution of above equation by simple calculation. Substituting a simple form of $c_{22}$, i.e. $c_{2}=\frac{ln(t^{16})}{16\alpha_1\beta_1}$, into equation (\ref{18}), infers equation (\ref{17}).

\begin{figure}[!htbp]
\centering
\subfigure[]{\includegraphics[height=4.2cm,width=4.2cm]{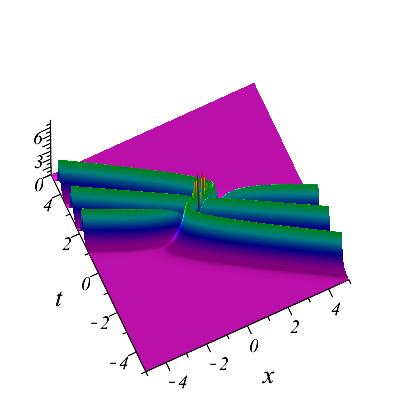}}
\subfigure[]{\includegraphics[height=4.2cm,width=4.2cm]{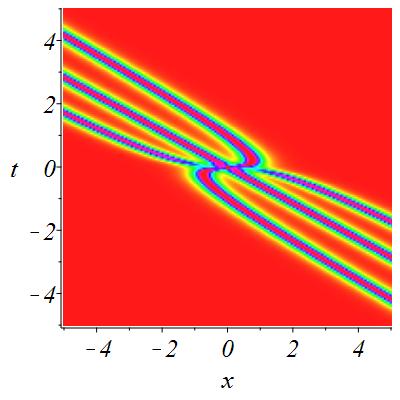}}
\subfigure[]{\includegraphics[height=4.2cm,width=4.2cm]{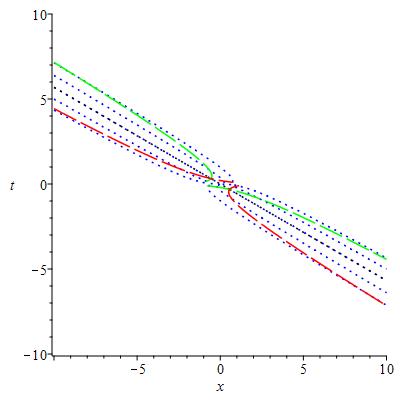}}
\caption{The evolution of three-positon $|q_{3-p}|^2$ with the parameter $\alpha_1=1.2$ and $\beta_1=1$ of the DNLS equation on $(x,t)$-plane. (a) is 3D plot, (b) is the density plot, (c) is the trajectory and the decomposition of three-positon solution with $c_{22}$. Five blue dots line in (c) for extreme values, two maxima denote the trajectories, but inner two minima is not. Two dashed line (green, upper; red, lower) denote approximate trajeceories. The middle black ($H=0$) which is the third approximate trajectory as the middle dot line denote trajectory without any ``phase shift".}\label{fig2}
\end{figure}

\noindent \textbf{Remark 3.2}: From the Proposition 3, it is easy to see the ``phase shift" is equivalent to $\ln(t^{16})$ when $t\rightarrow\infty$, differing from  the focusing mKdV equation (see Ref.\cite{xingqiuxia2017}) and the complex mKdV equation (see Ref.\cite{liuwei2017}). The ``phase shift" of the focusing mKdV equation and the complex mKdV equation is related with the $\ln({t^4})$. It is trivial to know that the ``phase shift" in Proposition 2 and Proposition 3 are equivalent to $\ln(t^4)$ are different from the ones of the focusing mKdV equation and the complex mKdV equation which are equivalent to $\ln(t^2)$ when $t\rightarrow\infty$. Furthermore, even $t$ is small, the more precise forms of the approximations about trajectories of positons is provied, i.e. ${c_{ij}}$,  there is very explicit expression in Fig. \ref{fig1}(c) and Fig. \ref{fig2}(c).

The trajectories of positon are given precisely in Figs. \ref{fig1}(c), \ref{fig2}(c), the line of the extreme maximum are plotted by blue dots. Moreover, in process of the decomposition, the approximate trajectories of positons are defined by $H\pm{c_{ij}}=0$ which are plotted in Figs. \ref{fig1}(c), \ref{fig2} (c) by green and red dashed lines. It's worth mentioning that the black line is plotted in Fig. \ref{fig2} (c) in the middle is also trajectory of the three-positon solution that is $H=0$. More importantly, the formulas of ``phase shift" ${c_{ij}}$ have been calculated explicitly in Proposition 2 and Proposition 3.

\begin{figure}[!htbp]
\centering
\subfigure[]{\includegraphics[height=4.2cm,width=4.2cm]{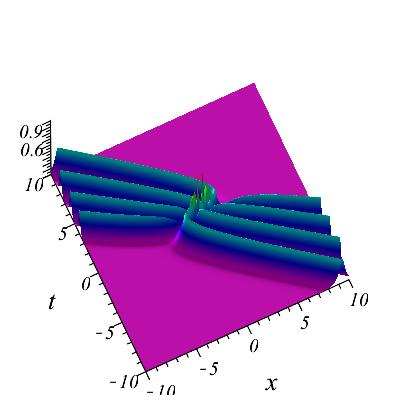}}
\subfigure[]{\includegraphics[height=4.2cm,width=4.2cm]{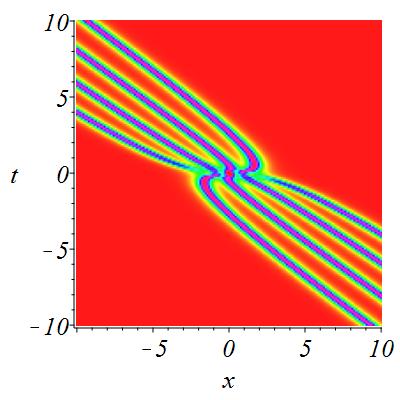}}
\caption{(a) is the evolution of the four-positon with the parameter $\alpha_1=1.2, \beta_1=1$ of the DNLS equation on $(x,t)$-plane, (b) is the density plots of solution.\label{fig3}}
\end{figure}

Generally, as for the larger $n$, the more complicated and interesting positon solution are got by similar method. In the next, because of the complexity of exact form of the four-positon soliton, we do not write down its explicit expression, but we have plotted it in Fig.\ref{fig3} and the higher order will not display because of intricacy.

\section{Combinations of solitons and positons}\label{section4}

In this section, we will discuss the hybrid of the soliton solutions and the positons solutions of DNLS equation. We omit the formulas which are indeed calculated complexly and the discussion about higher order mixed solutions of the DNLS equation.

\begin{figure}[!htbp]
\centering
\subfigure[]{\includegraphics[height=4.2cm,width=4.2cm]{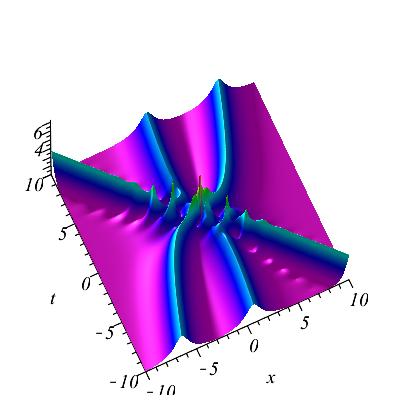}}
\subfigure[]{\includegraphics[height=4.2cm,width=4.2cm]{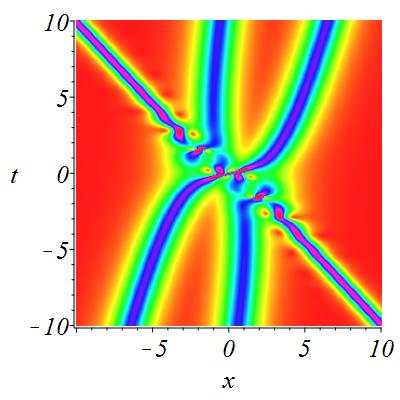}}
\caption{(a) is the combinations of one-soliton and two-positons with the parameter $\alpha_1=0.5,\beta_1=0.55,\alpha_5=0.85,\beta_5=0.7$ of the DNLS equation on $(x,t)$-plane. (b) is the density plots of solution.}\label{fig4}
\end{figure}

The $n$-positon solution is obtained after performing the higher-order Taylor expansion with $\lambda_j\rightarrow\lambda_1$ in the $N$-soliton solution. And two soliton can degenerate a positon, if some of $\lambda_j\rightarrow\lambda_1$ and others keep the original forms, namely not performing limit, then we can get mixed solutions about positons and solutions. Let's start from the lower order:

Case 1: $n=3$, we can get combination of one-soliton and two-positon solutions (see Fig.\ref{fig4}) with $\lambda_3\rightarrow\lambda_1$ and $\lambda_5$ remain unchanged.

\begin{figure}[!htbp]
\centering
\subfigure[]{\includegraphics[height=4.2cm,width=4.2cm]{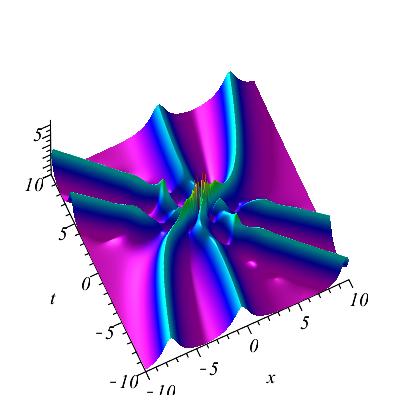}}
\subfigure[]{\includegraphics[height=4.2cm,width=4.2cm]{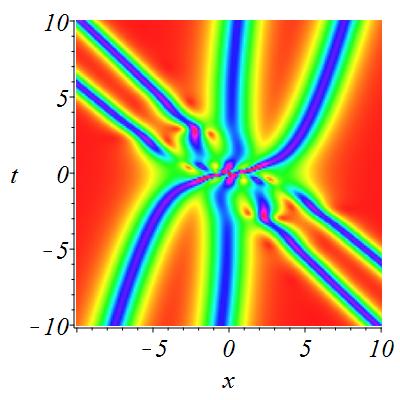}}
\caption{(a) is the combinations of two-soliton and two-positon with the parameter $\alpha_1=0.5, \beta_1=0.55, \alpha_5=0.8, \beta_5=0.6$ of the DNLS equation on $(x,t)$-plane. (b) is the density plots of solution.}\label{fig5}
\end{figure}

\begin{figure}[!htbp]
\centering
\subfigure[]{\includegraphics[height=4.2cm,width=4.2cm]{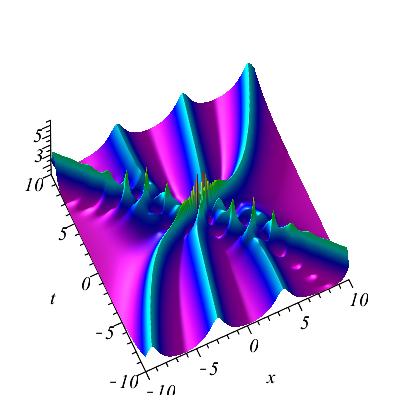}}
\subfigure[]{\includegraphics[height=4.2cm,width=4.2cm]{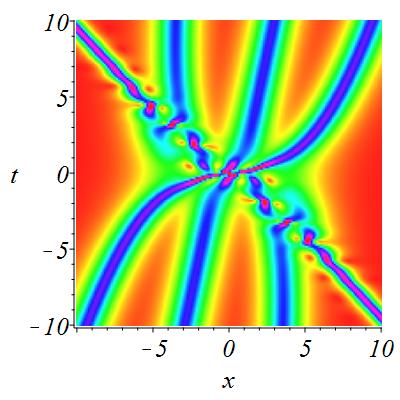}}
\caption{(a) is the combinations of one-soliton and three-positon with the parameter $\alpha_1=0.5, \beta_1=0.55, \alpha_7=0.85, \beta_7=0.7$ of the DNLS equation on $(x,t)$-plane. (b) is the density plots of solution.}\label{fig6}
\end{figure}

Case 2: $n=4$, the combinations of two-positon and two-positon solution (see Fig.\ref{fig5}) can be got with $\lambda_3\rightarrow\lambda_1$, $\lambda_7 \rightarrow\lambda_5$, but the mixed solutions of one-soliton and three-positon solution is obtained (see Fig.\ref{fig6}) with $\lambda_3\rightarrow\lambda_1$, $\lambda_5\rightarrow\lambda_1$ and $\lambda_7$ remain unchanged.

\noindent \textbf{Remark}: Fig. \ref{fig4}(a) display the hybrid of one-soliton and two-positon solutions and Fig. \ref{fig4}(b) is the density plots of solution, Fig. \ref{fig5}(a) display the hybrid of two-positon and two-positon solutions and Fig. \ref{fig5}(b) is the density plots of solution, Fig. \ref{fig6}(a) display the hybrid of one-soliton and three-positon solutions and Fig. \ref{fig6}(b) is the density plots of solution. From the picture, the main features of these new solutions are the following: the solutions are smooth and after the collision, two positons do not suffer a change of shape nor experience any asymptotic shift of phase. Asymptotically the soliton comes out of the collision with positon without any ``phase shift", but the positon gains two different phase shifts expressed in terms of the spectral parameters. The study of the solution will certainly enrich the theory of the DNLS equation. It is reasonable to suspect that the smooth mixed solutions of the positons and solitons of nature will likely exist for other nonlinear evolution equations as well.

\section{Conclusions}\label{section5}

From our study, the $n$-order smooth positon solutions of DNLS equation is provided explicitly by means of Taylor expansion in the corresponding determinant representation of the multi-soliton solution. Furthermore, we analyzed the crucial properties of positon solutions of DNLS equation from following three points of view: the decomposition, the approximate trajectories and the phase shift. From the Figs.\ref{fig1}, \ref{fig2}, \ref{fig3}, it is easy to see that the two-positon, three-positon, four-positon is not traveling wave, and the trajectory of positons is not a straight line, that is to say, it is a slowly changing curve.

It is worth mentioning that the ``phase shift" of the decomposition is different from focusing mKdV equation and complex mKdV equation. In the early stage of research on positons for complex mKdV equation, the distance of two peaks in two-positon of the complex mKdV equation is $2c_{11}$ ($\approx\frac{1}{2}\frac{\ln(2^{10}t^2\eta_1^6)}{\eta_1}$). As for focusing mKdV equation, the distance of two peaks in two-positon of the focusing mKdV equation is  $2c_{11}$ ($\approx\frac{\ln(64t^2)}{2}$). While the distance of two peaks in two-positon of the DNLS equation is $2c_{11}$ ($\approx\frac{\ln(8388608\alpha_1^8\beta_1^8(\alpha_1^2+\beta_1^2)t^4}{8\alpha_1\beta_1}$) which is different from them. In our paper, the ``phase shift" are equivalent to $\ln(t^4)$ when $t$ is very large, while the ``phase shift" of focusing mKdV equation and complex mKdV equation are equivalent to $\ln(t^2)$.  It is new finding and have never reported in literatures before.

 It is easy to see that the smooth $n$-positon soliton is a expression which is a mixture of polynomials and hyperbolic functions where is the similarity about the multi-pole solutions of the mKdV equation \cite{WMOK1982,N-double,exact,ekdv,focusing} and the NLS equation \cite{Olmedilla1989}, which were reported by means of the Hirota method and  the classical inverse scattering method in the past three decades. Compared our findings with the previous results, we come to the following conclusions: (1) the expression of $n$-positon solution is obtained simply and accurately in Eq.(\ref{npositon}); (2)it provide delicate and direct process for the decomposition of the modulus square in Eq.(\ref{13})  and  Eq. (\ref{17}), i.e. decomposing the multi-positon solutions into single solutions and the formulas of ``phase shift" and trajectories are shown precisely as well. The corresponding soliton trajectories are revealed clearly in Fig.\ref{fig1} and  Fig.\ref{fig2}.

We also discuss the combination of the solutions and positons. The interaction of positons and solitons can been seen from the Figs.\ref{fig4}, \ref{fig5} and \ref{fig6}. As for two positons, in the process of its interaction, the positons are not affected by the shape change or they do not suffer a change of any asymptotic shift of phase, which is a known characteristic to solitons. Positons are completely transparent to solitons, and vice versa, the positons are slightly altered by the solitons in a predictable way. During the soliton-positon collision, the soliton remains unchanged, while the carrier wave and the envelop of positon experience ``phase shift", which is keeping with the result of the Ref.\cite{Matveev19992,Chow1998}. A detailed analysis of the dynamical evolution of the degenerate solution of NLS has already been developed in Ref.\cite{He2013}. Wang \cite{Wang2017} et. al investigate a special kind of breather solution of the NLS equation, called breather-positon (or b-positon). The methods is similar to the means of the positons and worthy to study the b-positon solution of DNLS equation in the near future.

\noindent \textbf{Ethical Statement}\ {\small Authors declare that they comply with ethical standards. } \\
\noindent \textbf{Conflict of interest}\ {\small Authors declare that they have no conflict of interest.} \\
\noindent \textbf{Funding}\ This work is supported by
 the Natural Science Foundation of Zhejiang Province under Grant Nos. LY15A010005 and LZ19A010001, the Natural Science Foundation of Ningbo under Grant No. 2018A610197, the NSF of China under Grant Nos.11601187, 11671219, K.C.Wong Magna Fund in Ningbo University, Scientific Research Foundation of Graduate School of Ningbo University.

\end{document}